\def\om{\Omega}     
\def\omp{\Omega_{\rm p}}     
\def\oms{\Omega_{\rm s}}     
\def\omsa{\Omega_{\rm s,1}}     
\def\omsb{\Omega_{\rm s,2}}     
\def\kin{{\cal V}}     
\def\kinp{{\cal V}_{\rm p}}     
    
\def\pin{{\cal X}}     
\def\pinp{{\cal X}_{\rm p}}     
\def\pins{{\cal X}_{\rm s}}     
\def\vpd{{\cal R}}     
\def\vpdp{{\cal R}_{\rm p}}     
\def\vpds{{\cal R}_{\rm s}}     
     
\def\lenp{a_{\rm p}}     
\def\lens{a_{\rm s}}     
     
\def\lagp{D_{\rm p}}     
               
\def\kms{$\rm{km\;s^{-1}}$}     
\def\ma2{$\rm{mag\;arcsec^{-2}}$}

\documentstyle[12pt,aaspp4]{article} 

\begin{document}

\title{Direct Confirmation of Two Pattern Speeds in     
  the Double Barred Galaxy NGC 2950$^1$}
\footnotetext[1]{Based on observations made with 
the UK Jacobus Kapteyn Telescope and Italian Telescopio Nazionale
Galileo operated at the Spanish Observatorio del Roque de los
Muchachos of the Instituto de Astrof\'\i sica de Canarias by Isaac
Newton Group and Istituto Nazionale di Astrofisica, respectively.}
          
\author{E.M.~Corsini\altaffilmark{2},     
Victor P.~Debattista\altaffilmark{3}, and J.A.L.~Aguerri\altaffilmark{4}}     
\altaffiltext{2}{Dipartimento di Astronomia, Universit\`a di Padova,     
Vicolo dell'Osservatorio 2, I-35122 Padova, Italy, corsini@pd.astro.it}     
\altaffiltext{3}{Institut f\"ur Astronomie, ETH H\"onggerberg, HPF     
  G4.2, CH-8093 Z\"urich, Switzerland, debattis@phys.ethz.ch}     
\altaffiltext{4}{Instituto de Astrof\'{\i}sica de Canarias, V\'{\i}a     
  L\'actea s/n, E-38200 La Laguna, Spain, jalfonso@ll.iac.es}

\begin{abstract}     
     
  We present surface photometry and stellar kinematics of NGC 2950,
  which is a nearby and undisturbed SB0 galaxy hosting two nested
  stellar bars. We use the Tremaine-Weinberg method to measure the
  pattern speed of the primary bar. This also permits us to establish
  directly and for the first time that the two nested bars are
  rotating with different pattern speeds, and in particular that the
  rotation frequency of the secondary bar is higher than that of the
  primary one.
         
\end{abstract}


\section{Introduction}     
     
Large-scale bars are present in some $2/3$ of all disk galaxies
(Knapen et al 2000; Eskridge et al. 2000). Secondary stellar bars
within large-scale bars are also common, occurring in about $1/3$ of
barred galaxies (Laine et al. 2002; Erwin \& Sparke 2002).
Interest in secondary stellar bars is motivated by the hypothesis that
they are a mechanism for driving gas to small radii to feed the
supermassive black holes powering active galactic nuclei (e.g.  Regan
\& Mulchaey 1999).  However, the efficiency of such transport is
uncertain because of the complete lack of knowledge of $\omp$ and
$\oms$, the pattern speeds of the primary and secondary bars. Whereas
a number of pattern speeds of large-scale bars have been measured (see
Aguerri et al. 2003, hereafter ADC03 and references therein), no such
measurements in nested systems have been performed yet.
The presence of nested bars with different pattern speeds has been
inferred largely on the basis of their apparently random relative
orientations (Buta \& Crocker 1993).  The possibility that $\oms >
\omp$ is supported by the simulations of Rautiainen et al. (2002), 
which showed that secondary bars can naturally form and survive for
more than few rotation periods in pure stellar disks.  The
morphological characteristics of these systems are suggestive of stars
in the secondary bars oscillating about the loop orbits studied by
Maciejewski \& Sparke (2000) for models with $\oms >
\omp$. However, simulations have also found other possibilities (Friedli \&
Martinet 1993), including cases where two stellar bars counter-rotate
(Sellwood \& Merritt 1994; Friedli 1996).
     
A model-independent method for measuring pattern speeds is the
Tremaine-Weinberg method (Tremaine \& Weinberg 1984, TW hereafter).
This gives the pattern speed $\om$ of a single bar as
\begin{equation}     
\pin \om \sin i = \kin,      
\label{eq:tw1}     
\end{equation}     
where $\pin$ and $\kin$ are luminosity-weighted average position and
line-of-sight velocity measured parallel to the major axis of the
galaxy disk, and $i$ is the disk inclination. Long-slit spectra
parallel to the disk major-axis can measure all the quantities needed
by Eqn. \ref{eq:tw1} provided that the galaxy is free of extinction.
If several parallel slits at different offsets $Y$ relative to the
major axis are available for a galaxy, $\om \sin i$ can be obtained as
the slope of a plot of $\kin$ versus $\pin$.  In a double barred
galaxy (hereafter S2B), for a slit passing through both bars,
Eqn. \ref{eq:tw1} is modified to
\begin{equation}     
(\pinp \omp + \pins \oms) \sin i = \kin     
\label{eq:tw2}     
\end{equation}     
provided the two bars are rigidly rotating through each other, i.e.
that the total surface density is described by $\Sigma(R,\phi) =
\Sigma_p(R,\phi-\omp t) + \Sigma_s(R,\phi-\oms t)$. Eqn.
\ref{eq:tw2} is then a consequence of the linearity of the continuity
equation.  Eqn.  \ref{eq:tw2} can be solved for $\oms$ by first
measuring $\omp$ as in Eqn. \ref{eq:tw1} with slits which avoid the
secondary bar, and then modeling to obtain $\pins$ from the observed
$\pin = \pins + \pinp$.  However, the two bars are not likely to
rotate rigidly through each other when $\omp
\neq \oms$ (Louis \& Gerhard 1988; Maciejewski \& Sparke 2000;
Rautiainen et al.  2002), requiring additional modeling to obtain
$\oms$.  Nonetheless, when $\omp = \oms$, Eqn. \ref{eq:tw2} reduces to
Eqn. \ref{eq:tw1} and is satisfied exactly.  Thus testing whether
$\oms = \omp$ does not require any assumptions.

\section{NGC 2950}     
     
An ideal target for this purpose is the S2B NGC 2950, which is a large
($2\farcm7\times1\farcm8$ [de Vaucouleurs et al.  1991, hereafter
RC3]) and bright ($B_T=11.84$ [RC3]) early-type barred galaxy.  NGC
2950 is classified RSB0(r) and its total absolute magnitude is
$M_{B_T}^0=-20.03$ (RC3) adopting a distance of 23.3 Mpc (Tully 1988).
The presence of a secondary stellar bar has been discussed extensively
by Wozniak et al. (1995), Friedli et al.  (1996), and Erwin \& Sparke
(2002) on the basis of ground-based and {\it Hubble Space Telescope\/}
images in both optical and near-infrared bandpasses. The secondary bar
of NGC 2950 is typical of those in the samples of Laine et al. (2002)
and of Erwin \& Sparke (2002).
NGC 2950 meets all the requirements for the TW analysis: it has an
intermediate inclination, both bars have intermediate position angles
(hereafter PA) between the major and minor axes of the disk and no
evidence of spirals, patchy dust or significant companions.

\section{Surface photometry}     
     
The photometric observations of NGC 2950 were carried out at the 1-m  
Jacobus Kapteyn Telescope on December 27-28, 2000.  
We took multiple exposures in the Harris $B$ (4 $\times$ 1200 s), $V$
(3 $\times$ 480 s), and $I$ (18 $\times$ 150 s) bandpasses using the
SITe2 $2048\,\times\,2048$ CCD. This camera has a scale of $0\farcs33$
pixel$^{-1}$, yielding an unvignetted field of view of $\sim
10'\,\times\,10'$. The seeing FWHM was $\approx1\farcs0$.
The data reduction has been carried out using standard {\tt IRAF}
tasks as in Debattista et al. (2002, hereafter DCA02). Images were
bias subtracted, flatfielded, cleaned of cosmic rays, and corrected of
bad pixels. The sky-background level was removed by fitting a
second-order polynomial to the regions free of sources.
Photometric calibration, using standard stars, included corrections
for atmospheric and Galactic extinction, and for color as in DCA02. 
     
The radial profiles of surface brightness, ellipticity and PA were
obtained by fitting elliptical isophotes with the {\tt IRAF} task {\tt
ELLIPSE}. We first fitted ellipses allowing their centers to vary to
test for patchy dust obscuration. We found no evidence of a varying
center within the errors of the fits, and similar PA and ellipticity
profiles for all bandpasses. Thus we concluded that there is little or
uniform obscuration, as required for the TW method. The ellipse fits
were then repeated with the ellipse center fixed; the resulting
photometric profiles are plotted in Fig. 1.
We interpreted the local maximum in ellipticity at $r\simeq3''$ and
the corresponding twist and stationary value in PA as the photometric
signatures of the presence of a misaligned secondary bar inside the
primary bar. This is confirmed by the analysis of $B-I$ and $V-I$
color maps and unsharp mask of the original frames; in all we find no
evidence of other small-scale structures such as nuclear rings or
disks, spiral arms, star-forming regions, dust lanes and/or dust
patches, in agreement with previous results (Wozniak et al. 1995;
Friedli et al. 1996; Erwin \& Sparke 2003). In particular, the
structural details unveiled by the unsharp mask of the WFPC2/F814W
image of the nucleus of NGC 2950 (Erwin \& Sparke 2003) are unlike
those typical of nuclear stellar disks (Pizzella et al.  2002).
The PAs of the primary (${\rm PA_p}=152\fdg6\pm0\fdg4$) and secondary  
bar (${\rm PA_s}=91\fdg1\pm0\fdg3$) were measured in the $I-$band  
image at $r\simeq3''$ and $r\simeq23''$, at the two peaks in the  
ellipticity profile (Fig. 1).  
The lengths of the primary ($\lenp =34\farcs3\pm2\farcs1$) and
secondary bar ($\lens =4\farcs5\pm1\farcs0$) were measured using three
independent methods based on Fourier amplitudes (Aguerri et al. 2000),
Fourier and ellipse phases (DCA02), and a decomposition of the surface
brightness profiles (Prieto et al.  2001).
The inclination ($i=45\fdg6\pm1\fdg0$) and PA of the disk (${\rm  
PA_d}=116\fdg1\pm1\fdg0$) were determined by averaging the values  
measured between $65''$ and $100''$ in the $I-$band profile (Fig. 1).

\section{Long-slit spectroscopy}     
      
The spectroscopic observations of NGC 2950 were carried out at the  
3.6-m Te\-le\-sco\-pio Nazionale Galileo on December 18, 2001 (run 1),  
March 20-22, 2002 (run 2), and March 9-11, 2003 (run 3).  
The Low Resolution Spectrograph mounted the HR-V grism No.~6 with 600
$\rm grooves\;mm^{-1}$ and the $0\farcs7\,\times\,8\farcm1$ slit. The
detector was the Loral CCD with $2048\,\times\,2048$ pixels of
$15\,\times\,15$ $\mu$m$^2$. The wavelength range from 4660 to 6820
\AA\ was covered with a reciprocal dispersion of 1.054 \AA\
pixel$^{-1}$ and a spatial scale of 0.275 arcsec pixel$^{-1}$.
We obtained 4 spectra with the slit along the disk major axis (run 2
and 3), and 11 offset spectra with the slit parallel to it
($Y=-3\farcs1,+1\farcs5,+2\farcs8,\pm10\farcs1$ in run 1,
$Y=\pm5\farcs1,\pm13\farcs4$ in run 2, and $Y=-2\farcs5,+3\farcs5$ in
run 3, Fig. 2). The exposure times were $2\times60$ min and
$2\times45$ min for the major-axis spectra obtained in run 2 and 3,
respectively, and 45 min for all the offset spectra.  Comparison lamp
exposures before and/or after each object integration ensured accurate
wavelength calibrations. Spectra of G and K giant stars served as
kinematical templates. The seeing FWHM was $\approx1\farcs2$ in run 1,
$\approx1\farcs5$ in run 2, and $\approx0\farcs9$ in run 3.
Using standard MIDAS routines, all the spectra were bias subtracted,
flatfield corrected, cleaned of cosmic rays, corrected for bad pixels
and wavelength calibrated as in DCA02.  The accuracy of the wavelength
rebinning ($\lesssim2$ \kms) was checked by measuring wavelengths of
the brightest night-sky emission lines. The instrumental resolution
was 3.10 \AA\ (FWHM) corresponding to $\sigma_{\it inst} \approx 80$
\kms\ at 5170 \AA .
The major-axis spectra obtained in the same run were co-added using
the center of the stellar continuum as reference. In all the spectra
the contribution of the sky was determined by interpolating along the
outermost $\approx20''$ at the edges of the slit and then subtracted.

\section{Pattern speeds of the primary bar and secondary bar}     
     
To measure $\kin$ for each slit (Fig. 3a), we first collapsed each
two-dimensional spectrum along its spatial direction in the wavelength
range between 5060 and 5490 \AA, obtaining a one-dimensional spectrum.
The value of $\kin$ was then derived by fitting the resulting spectrum
with the convolution of the spectrum of the K1III star HR 4699 and a
Gaussian line-of-sight velocity profile by means of the Fourier
Correlation Quotient method (Bender 1990, hereafter FCQ) as done in
ADC03.  We estimated uncertainties by Monte Carlo simulations with
photon, read-out and sky noise.
To compute $\pin$ for each slit (Fig. 3b), we extracted the luminosity 
profiles from the $V-$band image along the position of the slit after 
convolving the image to the seeing of the spectrum. The $V-$band 
profiles match very well the profiles obtained by collapsing the 
spectra along the wavelength direction, confirming that the slits were 
placed as intended. We used the $V-$band profiles to compute $\pin$ 
because they are less noisy than those extracted from the spectra, 
particularly at large radii. 
We obtained $\omp$ from the values of $\pin$ and $\kin$ for the slits
at $|Y| \geq 3\farcs1$ and at $Y=0''$ (the latter constrain only the
zero point).  Since the slits at $|Y| \geq 3\farcs1$ do not cross the
secondary bar, we assume $\pin=\pinp$ and $\kin=\kinp$ for them and
obtain $\omp \sin{i}$ with a straight line fit. This gives $\omp= 11.2
\pm 2.4 $ \kms\ arcsec$^{-1}$ ($99.2 \pm 21.2 $ \kms\ kpc$^{-1}$, Fig. 3c)
The value of $\omp$ does not change within errors ($\omp= 10.9 \pm 2.4
$ \kms\ arcsec$^{-1}$) when we exclude the slits at
$Y=-3\farcs1,+3\farcs5$, confirming that they are not much affected by
the secondary bar (Fig. 2).
     
We used the FCQ to measure the line-of-sight velocity and velocity
dispersion profiles of the stellar component along the major axis
(Fig. 4). All the major-axis spectra were co-added after being
convolved to the same seeing.
We derived the circular velocity in the disk region, $V_c =
356^{+61}_{-49}$ \kms, after a standard correction for the asymmetric
drift as in ADC03.  Thus the corotation radius of the primary bar is
$\lagp = V_c/\omp = 32\farcs4_{-6.2}^{+8.7}$ and the ratio $\vpdp
\equiv \lagp/\lenp$ of the corotation radius to the bar semimajor axis 
is $\vpdp = 1.0_{-0.2}^{+0.3}$ (error intervals on $\lagp$ and $\vpdp$
are $68\%$ confidence level and were measured with Monte Carlo
simulations as in ADC03). We conclude that, within the errors, the
primary bar of NGC 2950 is consistent with all previous measurements
of $\vpd$ in SB0 galaxies (ADC03), which gives us confidence in our
assumption that the signals in the outer slits are generated by the
primary bar only.
     
The photometric and kinematic integrals measured with the innermost
slits ($|Y|\leq2\farcs8$) include a contribution from the secondary
bar. In particular, $|\kin| \gg |\kinp|$ for the slits at
$Y=-2\farcs5,+2\farcs8$ if we extrapolate $\kinp$ from large $|Y|$. A
straight line fit to $\pin$ and $\kin$ for the slits at
$|Y|\leq2\farcs8$ has a slope ($=63.7\pm7.1$ \kms\ arcsec$^{-1}$, Fig.
3c) which is different at better than $99\%$ confidence level from the
slope ($=8.0\pm1.7$ \kms\ arcsec$^{-1}$, Fig. 3c) of the straight line
fit for the primary bar. This may be because $\omp \neq \oms$;
however, to justify this conclusion, we needed to exclude the
possibility of a systematic error due to PA errors, which $\pin$ and
$\kin$ are quite sensitive to (Debattista 2003).
We tested whether the difference in the slopes of the straight line
fits may be due to only PA errors when $\omp = \oms$ by building a
model of a S2B galaxy with a single pattern speed from an $N-$body
simulation of a SB0 galaxy described in Debattista (2003).  We rotated
particles by 90$^\circ$, rescaled their phase space coordinates by a
factor of 0.2, and added them back to the original galaxy with various
primary to secondary mass ratios to approximately match NGC 2950.
After projecting this system as in NGC 2950, we proceeded to measure
$\pin$ and $\kin$ for slits misaligned with the major axis.  Even when
PA errors reached $\pm5^\circ$, we were not able to produce a system
which approaches the behavior of our observations.  In particular, we
were not able to produce a system in which the slopes of the integrals
plotted versus $|Y|$ are larger for $\kin$ and smaller for $\pin$ in
the region of the secondary bar than in the region of the main bar, as
we observed in NGC 2950 (Fig.  3a,b). We therefore concluded that the
signatures we observed could not be an artifact of any PA
misalignments on two bars rotating with a single pattern speed.  Our
results, therefore, lead us to conclude, directly and for the first
time, that $\oms \neq \omp$.
 
However, estimating $\oms$ is model dependent; we illustrate this by
considering two extreme cases.  First, we assumed that the secondary
bar dominates at $|Y| \leq 2\farcs8$, and used Eqn.
\ref{eq:tw1} with $\om$ replaced by $\oms$ to find $\omsa = 89.2 \pm
9.9$ \kms\ arcsec$^{-1}$ (Fig.  3c).  In the second case, we rewrote
Eqn.  \ref{eq:tw2} as 
$\pins (\oms - \omp) \sin{i}  = \kin - \omp \pin \sin{i}.$     
The observed quantities are $\pin$ and $\kin$, while $\omp$ was
measured above. To obtain $\pins$, first we derived the values of
$\pinp$ in the region of the nuclear bar by fitting a straight line to
the $\pin$ values at $|Y|\geq 10\farcs1$ (fits extending to smaller
$|Y|$ give larger $|\oms|$), and obtained $\pins = \pin - \pinp$.
Then plotting $(\kin - \omp \pin \sin{i})$ versus $\pins$, we obtain
$(\oms - \omp) \sin i$ as the slope of the best fitting line; the
result is ${\omsb} = -21.2\pm 6.2$ \kms\ arcsec$^{-1}$, {\it i.e.}
secondary bar counter-rotating relative to the primary.  (Note that
the range from $\omsa$ to $\omsb$ passes smoothly through $\pm
\infty$, {\it i.e.} a vertical line.)

\section{Conclusions}   
   
We showed that the primary bar in NGC 2950 is rapidly rotating.  If  
this is the norm in S2B galaxies, then it guarantees that primary bars  
are efficient at funnelling gas down to the radius of influence of  
secondary bars.    
In Fig. 4, we plot the lines of slope $\omsa \sin i$ and $\omsb \sin
i$.  The range of $\oms$ is large enough that it must include the case
where $\vpds \simeq 1$.  However, it also includes the case where
$\vpds \sim 2$, which hydrodynamical simulations find leads to
inefficient gas transport (Maciejewski et al.  2002).

We suggest two avenues for fruitful future work.  First, since the two
bars cannot be in exact solid body rotation (Louis \& Gerhard 1988;
Maciejewski \& Sparke 2000; Rautiainen et al. 2002), a more accurate
measurement of $\oms$ will require careful modeling and comparison
with simulations to account for such effects.  Second, it may be that
secondary bars oscillate about an orientation perpendicular to the
primary bar, possibly accounting for $\oms < 0$.  This can be tested
by repeating our measurements on a sample of S2B galaxies.
Nonetheless, we can confidently conclude that in NGC 2950 the two bars 
must have different pattern speeds, with the secondary bar having a 
larger pattern speed.

\newpage 

\begin{center}     
  \vspace*{15cm} \includegraphics{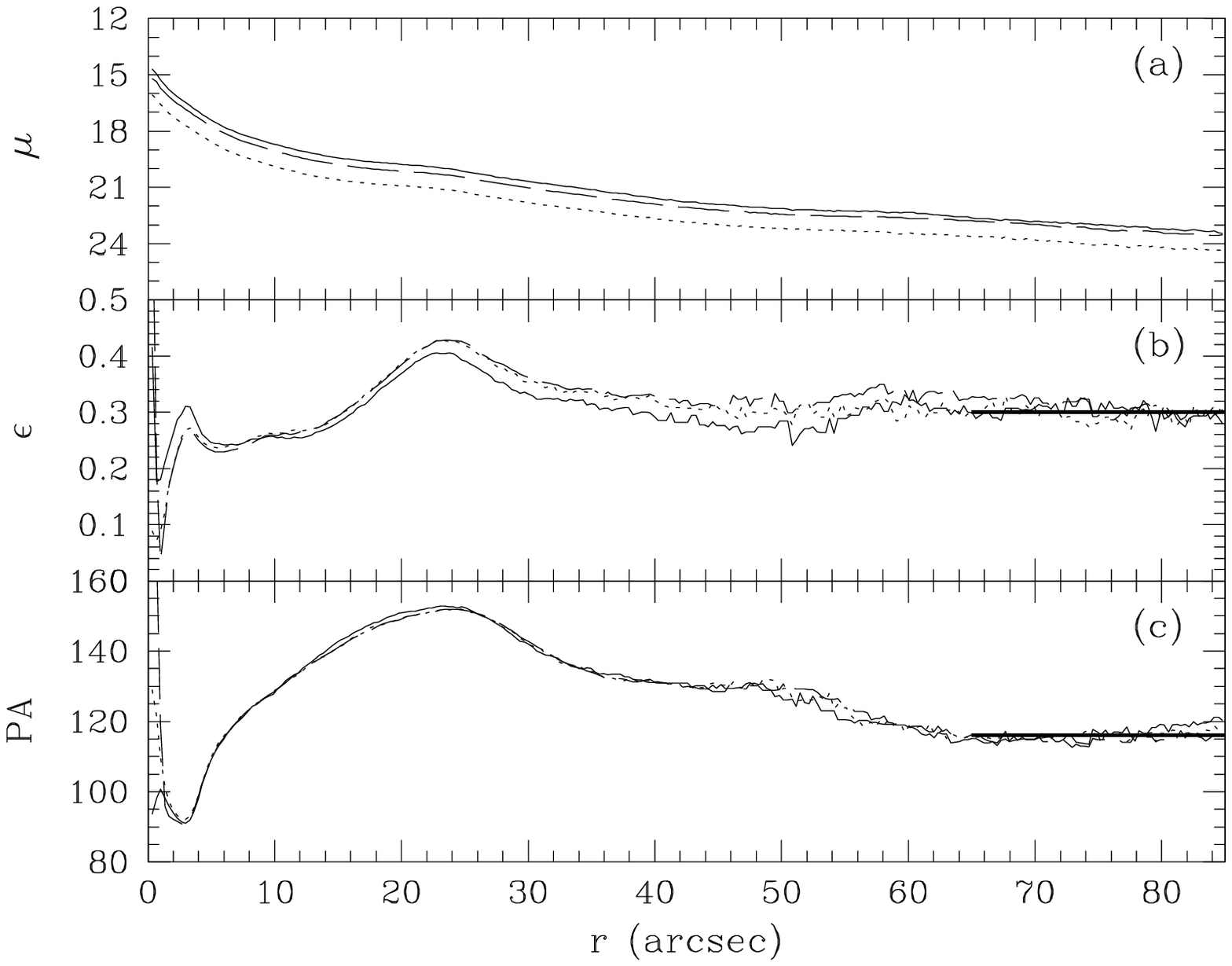}    
  \figcaption{ {\it (a)\/} Surface brightness, $\mu$, {\it (b)\/} 
    ellipticity, $\epsilon$, and {\it (c)\/} PA radial profiles of NGC 
    2950. Solid, dashed and dotted lines refer to $I$, $V$, and 
    $B-$band data, respectively. The thick lines represent the fits to 
    the $I-$band $\epsilon$ and PA of the galaxy's disk.} 
\end{center}

\newpage

\begin{center}     
  \vspace*{8cm} \includegraphics{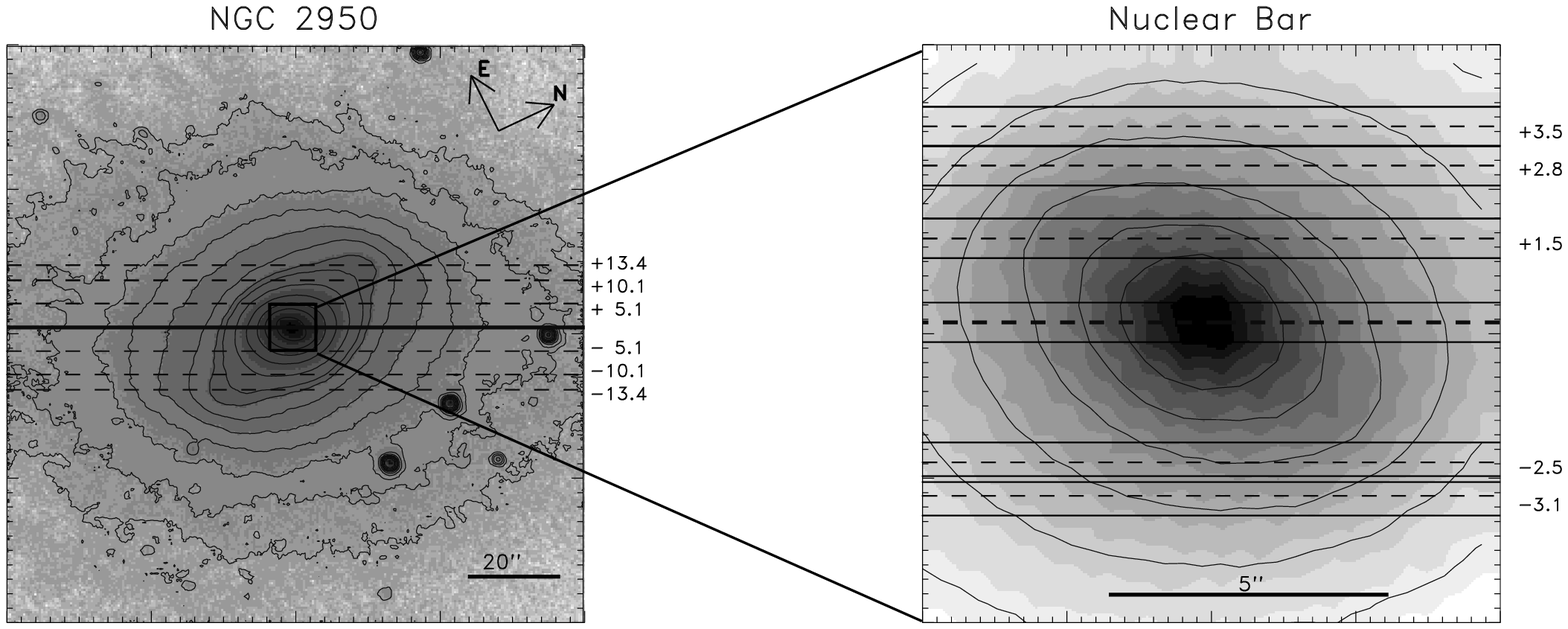}    
  \figcaption{{\it Left panel:} Large-scale image of NGC 2950 showing
    the primary bar and disk with $I-$band contours and slit positions
    overlaid.  Contours are spaced at 0.5 mag arcsec$^{-2}$ and the
    outermost corresponds to $\mu_I=23.0$ mag arcsec$^{-2}$. The solid
    and dashed lines correspond to the position of the spectra
    obtained along the disk major axis and at large offsets ($|Y| \geq
    5\farcs1$), respectively. For each slit position the offset, $Y$,
    is given in arcsec (we arbitrarily chose axes such that $Y$
    increases from the SW to the NE sides).
    {\it Right panel:} A zoom into the central region of NGC 2950
    showing its secondary bar. $I-$band isocontours are spaced at 0.5
    mag arcsec$^{-2}$, with the outermost corresponding to
    $\mu_I=19.5$ mag arcsec$^{-2}$. For each spectrum obtained at $|Y|
    \leq 3\farcs5$ the solid and dashed lines mark the edges and
    center of the slit, respectively.}
\end{center}

\newpage

\begin{center}     
  \vspace*{11cm} \includegraphics{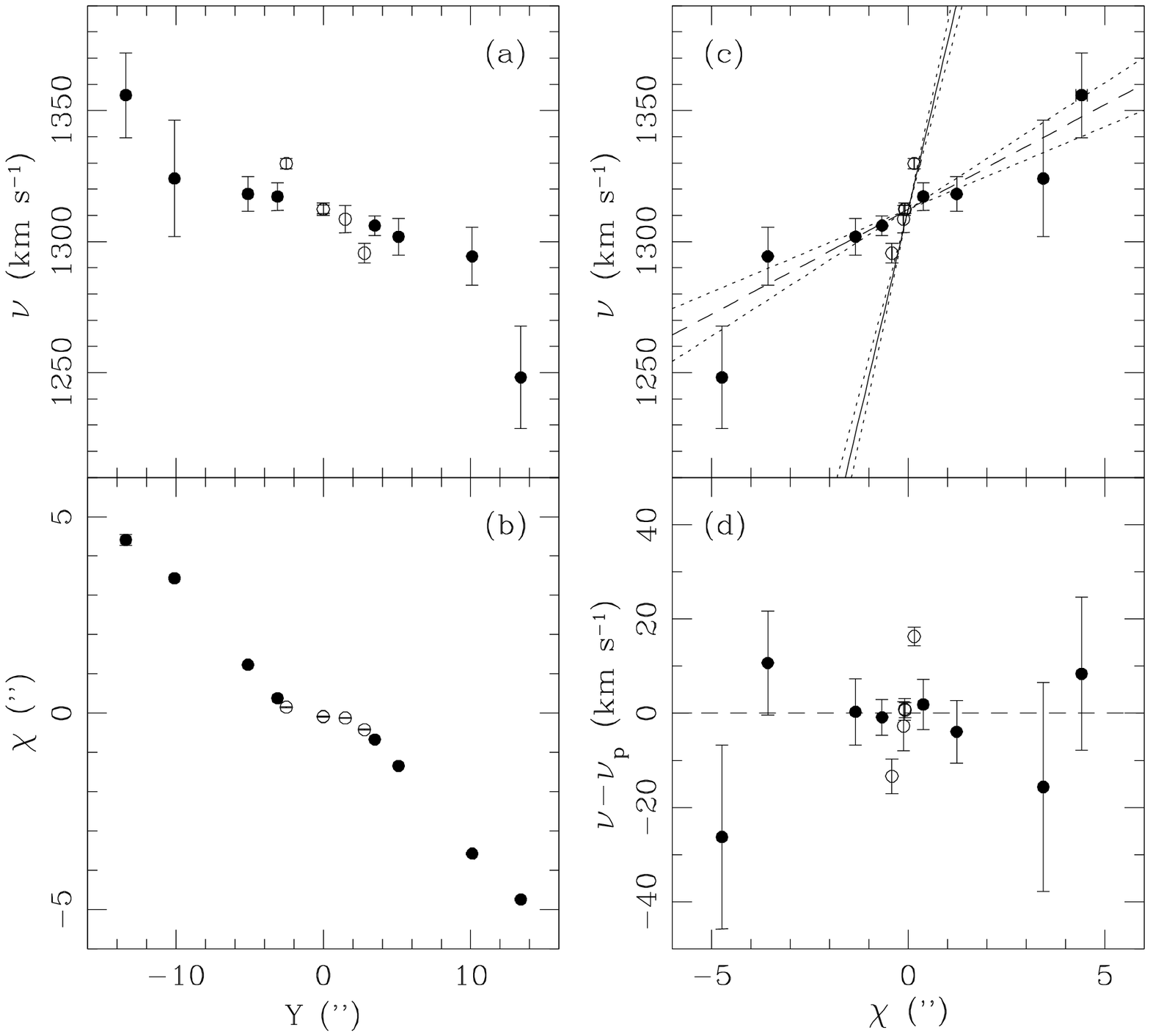}   
  \figcaption{{\it (a)\/} The kinematic integrals ${\cal V}$ as a
    function of the slit offset $Y$ with respect to the major axis
    ($Y=0''$).  Open and filled circles correspond to slits crossing
    the secondary bar ($|Y| \leq 2\farcs8$) and only the primary bar
    ($|Y| \geq 3\farcs1$), respectively.
    {\it (b)\/} The photometric integrals ${\cal X}$ as a function of   
    the slit offset $Y$.   
    {\it (c)} ${\cal V}$ as a function of ${\cal X}$ with different
    straight-line fits. These were obtained including the slits at
    $Y=0''$ and only the slits at $|Y|\geq3\farcs1$ (dashed line,
    slope $ \omp \sin{i} = 8.0\pm1.7$ \kms\ arcsec$^{-1}$) or the
    innermost slits at $|Y|\leq2\farcs8$ (solid line, slope $ \omsa
    \sin{i}= 63.7\pm7.1$ \kms\ arcsec$^{-1}$). The very different
    slopes of the two straight lines strongly suggest that the primary
    and secondary bars have different pattern speeds.
    {\it (d)} The residuals from the straight line fit to the slits at   
    $|Y|\geq3\farcs1$ and at $Y=0''$.}    
\end{center}

\newpage

\begin{center}     
  \vspace*{8cm} \includegraphics{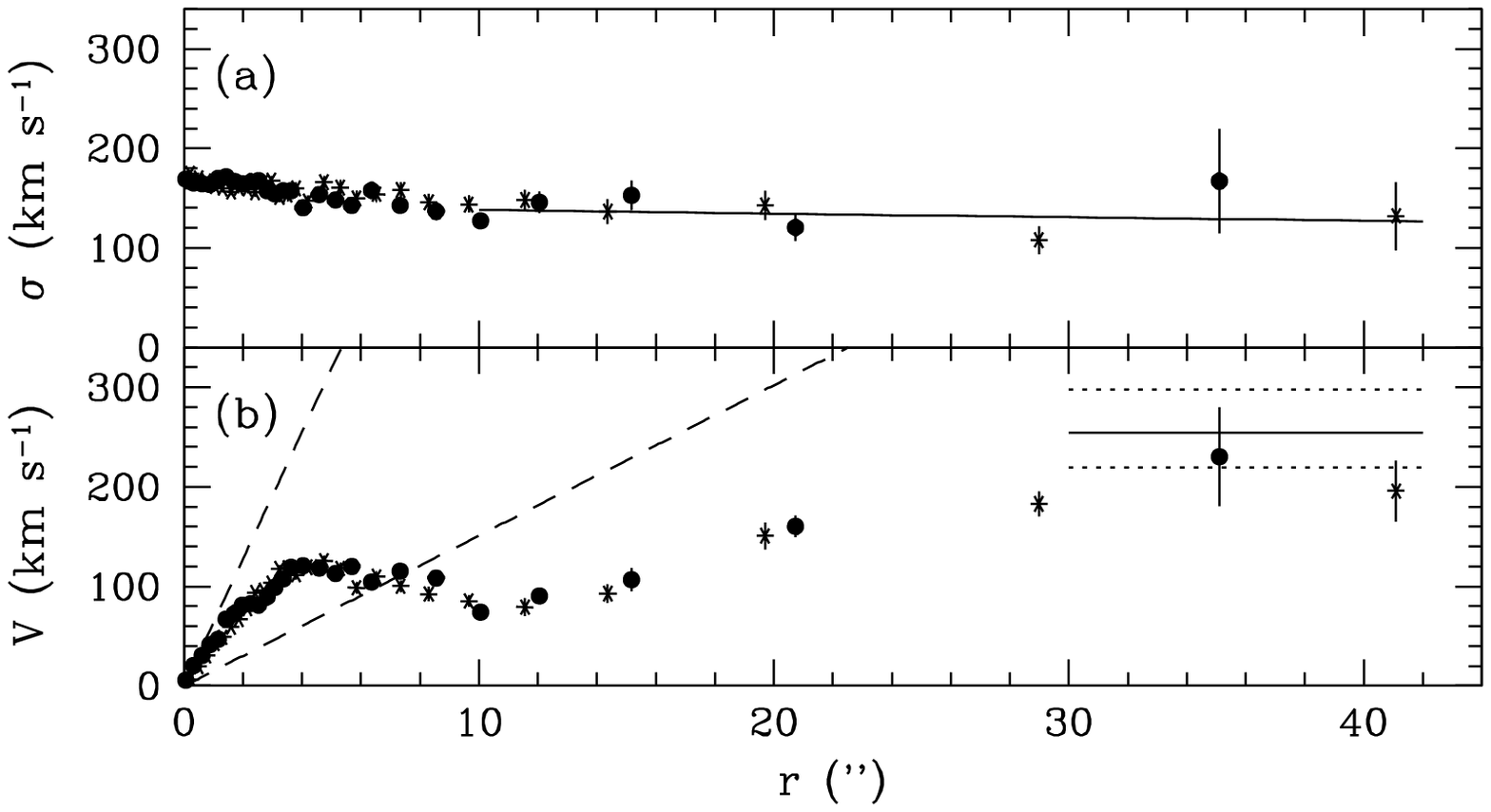}
  \figcaption{{\it (a)\/} The major-axis radial profile of the stellar
    line-of-sight velocity dispersion fitted with an exponential
    profile at $r\geq10''$ (solid line).
    {\it (b)\/} The major-axis radial profile of the stellar
    line-of-sight velocity (after subtracting the systemic velocity
    $V_{\it sys\/}=1312\pm3$ \kms) and the $V_c \sin{i}$ curve (solid
    line) with errors (dotted lines) obtained by applying the
    asymmetric drift for $r\geq30''$ as in ADC03. Dashed lines have
    slope $\omsa \sin i = 63.7$ \kms\ arcsec$^{-1}$ and $|\omsb|
    \sin{i} = 15.1$ \kms\ arcsec$^{-1}$.
    In {\it (a)\/} and {\it (b)\/} the measured profiles are folded
    around the center, with filled circles and asterisks referring to
    the SE (receding) and NW (approaching) sides, respectively.}
\end{center}

\end{document}